\documentclass[prl,amsmath,amssymb,twocolumn]{revtex4}

\usepackage{graphicx}

\newcommand{\be}{\begin{equation}}
\newcommand{\ee}{\end{equation}}
\newcommand{\ba}{\begin{eqnarray}}
\newcommand{\ea}{\end{eqnarray}}
\newcommand{\ban}{\begin{eqnarray*}}
\newcommand{\ean}{\end{eqnarray*}}

\newcommand{\moy}[1]{\langle #1 \rangle}

\newcommand{\ket}[1]{\mbox{$ | #1 \rangle $}}

\newcommand{\si}{\sigma}
\newcommand{\demi}{\frac{1}{2}}

\newcommand{\one}{\leavevmode\hbox{\small1\normalsize\kern-.33em1}}

\begin{document}

\title{Device-independent security of quantum cryptography against collective attacks}
\author{Antonio Ac\'{\i}n$^{1,2}$, Nicolas Brunner$^{3}$, Nicolas Gisin$^{3}$, Serge Massar$^{4}$,
Stefano Pironio$^{1}$, Valerio Scarani$^{3}$}
\address{$^1$ICFO-Institut de Ciencies Fotoniques, Mediterranean
Technology Park, 08860 Castelldefels (Barcelona), Spain\\
$^2$ICREA-Instituci\'o Catalana de Recerca i Estudis Avan\c{c}ats, 08010 Barcelona, Spain\\
$^3$Group of Applied Physics, University of Geneva, CH-1211 Geneve 4, Switzerland\\
%$^3$Group of Applied Physics, University of Geneva, 20 rue de l'Ecole-de-M\'edecine, CH-1211 Geneva 4, Switzerland\\
$^4$Laboratoire d'information quantique, CP 225, Universit\'e Libre
de Bruxelles, 1050 Brussels, Belgium }
%$^4$Laboratoire d'information quantique, Universit\'e Libre de Bruxelles, 50 Av. F.-D. Roosevelt, B-1050 Brussels, Belgium }
\date{June 25, 2007}

\begin{abstract}

We present the optimal collective attack on a Quantum Key
Distribution (QKD) protocol in the ``device-independent" security
scenario, where no assumptions are made about the way the QKD
devices work or on what quantum system they operate. Our main
result is a tight bound on the Holevo information between one of
the authorized parties and the eavesdropper, as a function of the
amount of violation of a Bell-type inequality.
\end{abstract}

%\pacs{PACS Nos. 03.65.Ud, 03.67.Dd}

\maketitle

Quantum Key Distribution (QKD) allows two parties, Alice and Bob,
to generate a secret key in the presence of an eavesdropper, Eve.
The secret key can then be used for different tasks, for instance
encryption of a message. In 1984, Bennett and Brassard invented
the first QKD protocol, known as BB84 \cite{bb84}. A few years
later, Ekert independently rediscovered that quantum correlations
can be used to distribute secrecy \cite{ekert}. Since then,
research on quantum cryptography has witnessed enormous advances,
both theoretical and experimental \cite{reviews}.

Existing QKD schemes rely for security on several assumptions. The
basic one is that \textit{any eavesdropper, however powerful, must
obey the laws of quantum physics}. In addition to it, there are
two other requirements, without which no shared secret key can be
established. The first one is the \textit{freedom and secrecy of
measurement settings}: on each particle, both Alice and Bob should
be allowed to choose freely among at least two measurement
settings (e.g., the two bases of BB84) and this choice should not
be known to Eve, at least as long as she can act on the incoming
quantum states (in BB84, the bases are revealed, but only after
the measurements are performed). The second requirement, even more
obvious, is the \textit{secrecy of outcomes}: at no stage there
should be a leakage of information about the final key. These two
requirements can be summarized by saying that \textit{no unwanted
classical information must leak out of Alice's and Bob's
laboratories}. If an implementation has a default in this point
(e.g., if a Trojan Horse attack is possible, or if Eve can access
Bob's computer), no security can be guaranteed.

In addition to these essential requirements, existing security
proofs \cite{SP,KGR,gllp} assume that Alice and Bob have (almost)
perfect control of the state preparation and of the measurement
devices. This assumption is often critical: for instance, the
security of the BB84 protocol is entirely compromised if Alice and
Bob, instead of sharing qubits as usually assumed, share
4-dimensional systems \cite{chshproto,mayers2}.

At first sight, control of the apparatuses seems to be an
inescapable requirement. Remarkably, this is not the case: we
present here a device-independent security proof against
collective attacks by a quantum Eve for the protocol described in
Ref.~\cite{amp06}. Our proof holds under no other requirements
than the essential ones listed above. It is therefore
``device-independent" in the sense that it needs no knowledge of
the way the QKD devices work, provided quantum physics is correct
and provided Alice and Bob do not allow any unwanted signal to
escape from their laboratories.

In a collective attack, Eve applies the same attack on each
particle of Alice and Bob, but no other limitations are imposed to
her. In particular she can keep her systems in a quantum memory
and perform a (coherent) measurement on them at any time.
Collective attacks are very meaningful in QKD because a bound on
the key rate for these attacks becomes automatically a bound for
the most general attacks if a de Finetti theorem can be applied,
as is the case in the usual security scenario \cite{Renner}.

The physical basis for our device-independent security proof is
the fact that measurements on entangled particles can provide
Alice and Bob with \textit{non-local correlations}, i.e.,
correlations that cannot be reproduced by shared randomness (local
variables), as detected by the violation of Bell-type
inequalities. Considered in the perspective of QKD, the fact that
Alice's and Bob's symbols are correlated in a non-local way,
whatever be the underlying physical details of the apparatuses
that produced those symbols, implies that Eve cannot have full
information about them, otherwise her own symbol would be a local
variable able to reproduce the correlations.

This intuition has been around for some time
\cite{ekert,mayers,BBM}. Quantitative progress has been possible
however only recently, thanks to the pioneering work of Barrett,
Hardy and Kent \cite{bhk} and to further extensions
\cite{chshproto,amp06,masanes}. For conceptual interest and
mathematical simplicity, all these works studied security against
a supra-quantum Eve, who could perform any operation compatible
with the no-signalling principle. The proof of Ref.~\cite{bhk}
applies only to the zero-error case; those in Refs
\cite{chshproto,amp06} allow for errors but restrict Eve to
perform individual attacks; Masanes and Winter \cite{masanes}
proved non-universally-composable security under the assumption
that Eve's attack is arbitrary but is not correlated with the
classical post-processing of the raw key. In this paper, we focus
on the more realistic situation in which Eve is constrained by
quantum physics, and prove universally-composable security against
collective attacks.

\textit{The protocol}. The protocol that we study is a
modification of the Ekert 1992 protocol \cite{ekert} proposed in
Ref.~\cite{amp06}. Alice and Bob share a quantum channel
consisting of a source that emits pairs of entangled particles. On
each of her particles Alice chooses between three possible
measurements $A_0$, $A_1$ and $A_2$, and Bob between two possible
measurements $B_1$ and $B_2$. All measurements have binary
outcomes labelled by $a_i,b_j\in\{+1,-1\}$ (note however that the
quantum systems may be of dimension larger than 2). The raw key is
extracted from the pair $\{A_0,B_1\}$. In particular, the quantum
bit error rate (QBER) is $Q=\text{prob}(a_0\neq b_1)$. As
mentioned in the introduction, Eve's information is bounded by
evaluating Bell-type inequalities, since these are the only
entanglement witnesses which are independent of the details of the
system. In our case, Alice and Bob use the measurements $A_{1}$,
$A_2$, $B_1$, and $B_2$ on a subset of their particles to compute
the Clauser-Horne-Shimony-Holt (CHSH) polynomial \cite{chsh} \be
{\cal S}=\moy{a_1b_1}+\moy{a_1b_2}+ \moy{a_2b_1}- \moy{a_2b_2}\,,
\ee which defines the CHSH inequality ${\cal S}\leq 2$.  We note
that there is no a priori relation between the value of ${\cal S}$
and the value of $Q$: these are the two parameters which are
available to estimate Eve's information. Without loss of
generality, we suppose that the marginals are random for each
measurement, i.e., $\moy{a_i}=\moy{b_j}=0$ for all $i$ and $j$.
Were this not the case, Alice and Bob could achieve it a
posteriori through public one-way communication by agreeing on
flipping a chosen half of their bits. This operation would not
change the value of $Q$ and $\mathcal{S}$ and would be known to
Eve.

\textit{Eavesdropping.} In the device-independent scenario, Eve is
assumed not only to control the source (as in usual
entanglement-based QKD), but also to have fabricated Alice's and
Bob's measuring devices. The only data available to Alice and Bob to
bound Eve's knowledge are the observed relation between the
measurement settings and outcomes, without any assumption on how the
measurements are actually carried out or on what system they
operate. In complete generality, we may describe this situation as
follows. Alice, Bob, and Eve share a state $\ket{\Psi}_{ABE}$ in
$\mathcal{H}_A^{\otimes n}\otimes \mathcal{H}_B^{\otimes n} \otimes
\mathcal{H}_E$, where $n$ is the number of bits of the raw key. The
dimension $d$ of Alice and Bob Hilbert spaces
$\mathcal{H}_A=\mathcal{H}_B=\mathbb{C}^d$ is unknown to them and
fixed by Eve. The measurement $M_k$ yielding the $k^\mathrm{th}$
outcome of Alice is defined on the $k^\mathrm{th}$ subspace of Alice
and chosen by Eve. This measurement depends on the $k^\textrm{th}$
setting $A_{j_k}$ chosen by Alice, but possibly also on all previous
settings and outcomes:
$M_k=M(A_{j_k},\bar{A}_{k-1},\overline{a}_{k-1})$ where
$\bar{A}_{k-1}=(A_{j_1},\ldots,A_{j_{k-1}})$ and
$\bar{a}_{k-1}=(a_{j_1},\ldots,a_{j_{k-1}})$. The situation is
similar for Bob.

\textit{Collective attacks.} In this paper, we focus on collective
attacks where Eve applies the same attack to each system of Alice
and Bob. Specifically, we assume that the total state shared by
the three parties has the product form
$\ket{\Psi_{ABE}}=\ket{\psi_{ABE}}^{\otimes n}$ and that the
measurements are a function of the current setting only, e.g., for
Alice $M_k=M(A_{j_k})$. (From now on, we thus simply write the
measurement $M(A_j)$ as $A_j$).

For collective attacks, the secret key rate $r$ under one-way
classical postprocessing from Bob to Alice is lower-bounded by the
Devetak-Winter rate \cite{DW}, \be \label{rate} r\, \geq\,
r_{DW}\,= \,I(A_0:B_1)\,-\, \chi(B_1:E)\,, \ee which is the
difference bewteen the mutual information between Alice and Bob,
$I(A_0:B_1)=1-h(Q)$ ($h$ is the binary entropy), and the Holevo
quantity between Eve and Bob, $\chi(B_1:E)=S(\rho_E)-
\demi\sum_{b_1=\pm 1}S(\rho_{E|b_1}) $. Note that the rate is
given by \eqref{rate} because $\chi(A_0:E)\geq \chi(B_1:E)$ holds
for our protocol \cite{amp06,full}; it is therefore advantageous
for Alice and Bob to do the classical postprocessing with public
communication from Bob to Alice.

\textit{Upper-bound on the Holevo quantity.} To find Eve's optimal
collective attack, we must find the largest value of $\chi(B_1:E)$
compatible with the observed parameters without assuming anything
about the physical systems and the measurements that are
performed. Our main result is the following.

\textit{Theorem.} Let $\ket{\psi_{ABE}}$ be a quantum state and
$\{A_1,A_2,B_1,B_2\}$ a set of measurements yielding a violation
$\mathcal{S}$ of the CHSH inequality. Then after Alice and Bob
have symmetrized their marginals, \be \chi(B_1:E)\leq
h\left(\frac{1+\sqrt{({\cal S}/2)^2-1}}{2}\right)\,. \label{bound}
\ee

Before presenting the proof of this bound, we give an explicit
attack which saturates it; this example clarifies why the bound
(\ref{bound}) is independent of $Q$. Eve sends to Alice and Bob
the two-qubit Bell-diagonal state \be \rho_{AB}({\cal
S})=\frac{1+{\cal C}}{2}\,P_{\Phi^+}\,+\, \frac{1-{\cal
C}}{2}\,P_{\Phi^-}\,, \label{rhoabc} \ee where $P_{\Phi^\pm}$ are
the projectors on the Bell states
$\ket{\Phi^\pm}=(\ket{00}\pm\ket{11})\sqrt{2}$ and ${\cal
C}=\sqrt{({\cal S}/2)^2-1}$. She defines the measurements to be
$B_1=\si_z$, $B_2=\si_x$ and $A_{1,2}=\frac{1}{\sqrt{1+{\cal
C}^2}}\si_z\pm\frac{{\cal C}}{\sqrt{1+{\cal C}^2}}\si_x$. Any
value of $Q$ can be obtained by choosing $A_0$ to be $\si_z$ with
probability $1-2Q$ and to be a randomly chosen bit with
probability $2Q$. This attack is impossible within the usual
assumptions because here not only the state $\rho_{AB}$, but also
the measurements taking place in Alice's apparatus depend
explicitly on the observed values of ${\cal S}$ and $Q$. The state
(\ref{rhoabc}) has a nice interpretation: it is the two-qubit
state which gives the highest violation ${\cal S}$ of the CHSH
inequality for a given value of the entanglement, measured by the
concurrence ${\cal C}$ \cite{vw}.

We now present the proof of the Theorem stated above, in four
steps; see Ref.~\cite{full} for more details.

\textit{Proof, Step 1.} It is not restrictive to suppose that Eve
sends to Alice and Bob a mixture $\rho_{AB}=\sum_c
p_c\,\rho_{AB}^c$ of two-qubit states, together with a classical
ancilla (known to her) that carries the value $c$ and determines
which measurements $A_{i}^c$ and $B_{j}^c$ are to be used on
$\rho_{AB}^c$.

The proof of this first statement relies critically on the
simplicity of the CHSH inequality (two binary settings on each
side). We present the argument for Alice, the same holds for Bob.
First, we may assume that the two measurements $A_{1,2}$ of Alice
are von Neumann measurements, if necessary by including ancillas
in the state $\rho_{AB}$ shared by Alice and Bob. Thus $A_1$ and
$A_2$ are hermitian operators on $\mathbb{C}^d$ with eigenvalues
$\pm 1$. It follows from this that $A_1A_2$ is a unitary, hence
diagonalizable, operator. In the basis of $\mathbb{C}^d$ formed by
the eigenvectors of $A_1A_2$, one can show that $A_1$ and $A_2$
are block-diagonal, with blocks of size $1\times 1$ or $2\times 2$
\cite{full,lluis}. In other words, $A_j=\sum_c P_c A_j P_c$ where
the $P_c$s are projectors of rank $1$ or $2$. From Alice's
standpoint, the measurement of $A_i$ thus amounts at projecting in
one of the (at most) two-dimensional subspaces defined by the
projectors $P_c$, followed by a measurement of the reduced
observable $P_c A_i P_c=\vec{a}\,^c_i\cdot\vec{\sigma}$. Clearly,
it cannot be worse for Eve to perform the projection herself
before sending the state to Alice and learn the value of $c$. The
same holds for Bob. We conclude that in each run of the experiment
Alice and Bob receive a two-qubit state. The deviation from usual
proofs lies in the fact that the measurements to be applied can
depend explicitly on the state.

\textit{Proof, Step 2.} Each state $\rho_{AB}^c$ can be taken to
be a Bell-diagonal state and the measurements $A_i^c$ and $B_j^c$
to be measurements in the $(x,z)$ plane.

To reduce the problem further in this way, we use some freedom in
the labeling together with two applications of a usual argument.
For fixed $c$ (we now omit the index $c$), let us first choose the
axis of the Bloch sphere on Alice's side in such a way that
$\vec{a}_1$ and $\vec{a}_2$ define the $(x,z)$ plane, and
similarly on Bob's side. Eve is a priori distributing any
two-qubit state $\rho$ of which she holds a purification. Now,
recall that we have supposed, without loss of generality, that all
the marginals are uniformly random. Here comes an argument which
is typical of QKD \cite{KGR}: knowing that Alice and Bob are going
to symmetrize their marginals, Eve does not lose anything in
providing them a state with the suitable symmetry. The reason is
as follows. First note that since the (classical) randomization
protocol that ensures $\moy{a_i}=\moy{b_j}=0$ is done by Alice and
Bob through public communication, we can as well assume that it is
Eve who does it, i.e., she flips the value of each outcome bit
with probability one half. But because the measurements of Alice
and Bob are in the $(x,z)$ plane, we can equivalently, i.e.,
without changing Eve's information, view the classical flipping of
the outcomes as the quantum operation $\rho\rightarrow
\tilde{\rho}=(\sigma_y\otimes\sigma_y)
\rho(\sigma_y\otimes\sigma_y)$ on the state $\rho$. We conclude
that it is not restrictive to assume that Eve is in fact sending
the mixture $\bar{\rho}=\demi\left(\rho+\tilde{\rho}\right)$,
i.e., that she is sending a state invariant under
$\si_y\otimes\si_y$. Now, through an appropriate choice of basis
that leaves invariant the ($x,z)$ plane, and corresponding to the
freedom to define the orientation of $\hat{y}$ and the direction
of $\hat{x}$ for both Alice and Bob (see \cite{full} for the
explicit transformations), every $\si_y\otimes\si_y$ invariant
two-qubit state can be written in the Bell basis, ordered as
$\{\ket{\Phi^+},\ket{\Psi^-}, \ket{\Phi^-},\ket{\Psi^+}\}$, in the
canonical form \ba \bar{\rho}&=&\left(\begin{array}{cccc}
\lambda_{\Phi^+} &
ir_1\\
-ir_1& \lambda_{\Psi^-}\\ && \lambda_{\Phi^-} & ir_2\\ && -ir_2&
\lambda_{\Psi^+}
\end{array}\right)\,,
\ea with $\lambda_{\Phi^+}\geq \lambda_{\Psi^-}$,
$\lambda_{\Phi^+}\geq \lambda_{\Phi^-}\geq\lambda_{\Psi^+}$ and
$r_1$, $r_2$ real.

Finally, we repeat an argument similar to the one given above:
since $\bar{\rho}$ and its conjugate $\bar{\rho}^{\,*}$ produce
the same statistics for Alice and Bob's measurements and provide
Eve with the same information, we can suppose without loss of
generality that Alice and Bob rather receive the mixture $\demi
\left(\bar{\rho}+\bar{\rho}^{\,*}\right)$, which is Bell-diagonal.

\textit{Proof, Step 3.} For a Bell-diagonal state $\rho_{\lambda}$
with eigenvalues ${\lambda}$ ordered as above and for measurements
in the $(x,z)$ plane, \be \chi_{{\lambda}}(B_1:E)\leq F({\cal
S}_{{\lambda}})= h\left(\frac{1+\sqrt{({\cal
S}_{{\lambda}}/2)^2-1}}{2}\right)\,, \label{estimate2} \ee where
$\mathcal{S}_{{\lambda}}=2\sqrt{2}\sqrt{(\lambda_{\Phi^+}-\lambda_{\Psi^-})^2+(\lambda_{\Phi^-}-
\lambda_{\Psi^+})^2}$ is the largest violation of the CHSH
inequality by the state $\rho_{\lambda}$.

This step is mainly computational; we sketch it here and refer to
\cite{full} for details. For Bell-diagonal states, for any choice
of $B_1=\cos\varphi\si_z+\sin\varphi\si_x$, one has
$S(\rho_{E|b_1=0})=S(\rho_{E|b_1=1})\geq h(\lambda_{\Phi^+}+
\lambda_{\Phi^-})$ with equality if and only if $B_1=\sigma_z$. It
follows that \be \chi_\lambda(B_1:E)\leq
H({\lambda})-h(\lambda_{\Phi^+}+ \lambda_{\Phi^-})\,.
\label{estimate1} \ee The right hand side of this expression is in
turn bounded by the function $F(\mathcal{S}_\lambda)$ appearing in
\eqref{estimate2}. It now suffices to notice that
$\mathcal{S}_\lambda= 2\sqrt{2}\sqrt{(\lambda_{\Phi^+}-
\lambda_{\Psi^-})^2+ (\lambda_{\Phi^-}- \lambda_{\Psi^+})^2}$ is
the maximal violation of the CHSH inequality by the  state
$\rho_{\lambda}$ \cite{vw,horodecki}; it is achieved for
$B_1=\si_z$, $B_2=\si_x$, and $A_{1}$ and $A_2$ depending
explicitly on the $\lambda$'s.

\textit{Proof, Step 4.} To conclude the proof, note that if Eve
sends a mixture of Bell-diagonal states $\sum_{\lambda}
p_\lambda\,\rho_{\lambda}$ and chooses the measurements to be in
the $(x,z)$ plane, then $\chi(B_1:E)=\sum_\lambda p_\lambda\,
\chi_{\lambda}(B_1:E)$. Using \eqref{estimate2}, we then find
$\chi(B_1:E)\leq \sum_\lambda p_\lambda\,
F(\mathcal{S}_\lambda)\leq F(\sum_\lambda p_\lambda\,
\mathcal{S}_\lambda)$, where the last inequality holds because $F$
is concave. But since the observed violation $\mathcal{S}$ of CHSH
is necessarily such that $\mathcal{S}\leq \sum_\lambda p_\lambda
\mathcal{S}_\lambda$ and since $F$ is a monotonically decreasing
function, we find $\chi(B_1:E)\leq F(\mathcal{S})$.

\begin{figure}
\includegraphics[scale=0.55]{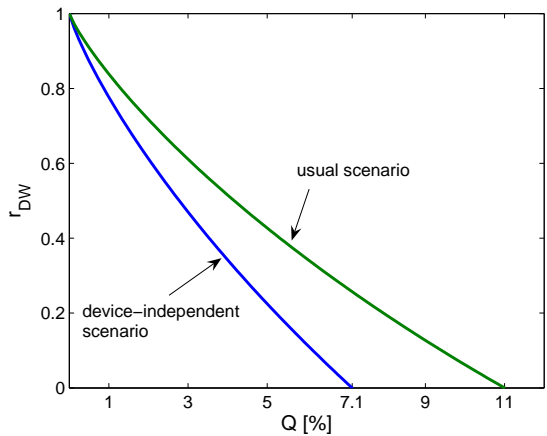}
\caption{Extractable secret-key rate against collective attacks in
the usual scenario [$\chi(B_1:E)$ given by
eq.~(\ref{boundstandard})] and in the device-independent scenario
[$\chi(B_1:E)$ given by eq.~(\ref{bound})], for correlations
satisfying $\mathcal{S}=2\sqrt{2}(1-2{Q})$.} \label{figcurves}
\end{figure}

\textit{Key rate.} Given the bound (\ref{bound}), the key rate
\eqref{rate} can be computed for any values of ${Q}$ and
$\mathcal{S}$. As an illustration, we study correlations
satisfying $\mathcal{S}=2\sqrt{2}(1-2{Q})$, and which arise from
the state $\ket{\Phi^+}$ after going through a depolarizing
channel, or through a phase-covariant cloner, or more generally
from any Bell-diagonal $\rho_{AB}$ such that $\lambda_{\Phi^+}\geq
\lambda_{\Psi^-}$ and $\lambda_{\Phi^-}=\lambda_{\Psi^+}$, when
doing the measurements $A_0=B_1=\sigma_z$, $B_2=\sigma_x$,
$A_1=(\si_z+\si_x)/{\sqrt{2}}$ and $A_2=(\si_z-\si_x)/\sqrt{2}$.
We consider these correlations because of their experimental
significance, but it is important to stress that Alice and Bob do
not need to assume that they perform the above qubit measurements.
The corresponding key rate is plotted in Fig.~\ref{figcurves} as a
function of $Q$. For the sake of comparison, we have also plotted
the key rate under the usual assumptions of QKD for the same set
of correlations. In this case, Alice and Bob have a perfect
control of their apparatuses, which we have assumed to faithfully
perform the qubit measurements given above. The protocol is then
equivalent to Ekert's, which in turn is equivalent to the
entanglement-based version of BB84: Eve's information on the key
is determined by the ``phase error'' \cite{SP}, which can be
computed for our protocol using the formalism of Ref.~\cite{KGR}.
One finds $e_p=1-Q-{\cal S}/2\sqrt{2}$, whence \be \chi(B_1:E)\leq
h\left(Q+{\cal S}/2\sqrt{2}\right)\,. \label{boundstandard} \ee If
$e_p=Q$, i.e., $\mathcal{S}=2\sqrt{2}(1-2{Q})$, this expression
yields the well-known critical QBER of $11\%$ \cite{SP}, to be
compared to $7.1\%$ in the device-independent scenario
(Fig.~\ref{figcurves}). (Note that the key rate given by eq.~\eqref{bound} is much higher than the one against a no-signalling eavesdropper obtained by applying the security proof of \cite{masanes}.)

\textit{Final remarks.} Through its remarkable generality, our
device-independent security proof allows us to ignore the detailed
implementation of the QKD protocol and therefore applies in a simple
way to situations where the quantum apparatuses are noisy or where
uncontrolled side channels are present. It also applies to the
situation where the apparatuses are entirely untrusted and provided
by the eavesdropper herself. In this latter case, the proof cannot
be applied to any existing device yet, because of the detection
loophole which arises due to inefficient detectors and photon
absorption. These processes imply that sometimes Alice's and Bob's
detectors will not fire. A possible strategy to apply our proof to
this new situation is for Alice and Bob to replace the absence of a
click by a chosen outcome, in effect replacing detection
inefficiency by noise. However the amount of detection inefficiency
that can be tolerated in this way is much lower than the one present
in current quantum communication experiments. In Bell tests, this
problem is often circumvented by invoking additional assumptions
such as the fair sampling hypothesis --- a very reasonable one if
the aim is to constrain possible models of Nature, but hardly
justified if the device is provided by an untrusted Eve. In the
light of the present work, the ``detection loophole" thus becomes a
meaningful issue in applied physics.

In conclusion, we have found the optimal collective attack on a
QKD protocol in the device-independent scenario, in which no other
assumptions are made than the validity of quantum physics and the
absence of any leakage of classical information from Alice's and
Bob's laboratories. If a suitable de Finetti-like theorem can be
demonstrated in this scenario, the bound that we have presented
here will in fact be the bound against the most general attacks.

\textit{Acknowledgements.} We are grateful to C.~Branciard,
I.~Cirac, A.~Ekert, A.~Kent, R.~Renner and C.~Simon for fruitful
discussions. We acknowledge financial support from the Swiss NCCR
``Quantum Photonics", the EU Qubit Applications Project (QAP)
Contract number 015848, the Spanish projects FIS2004-05639-C02-02
and Consolider QOIT, the Spanish MEC for a ``Juan de la Cierva"
grant, and the IAP project Photonics@be of the Belgian Science
Policy.

\end{document}